\documentclass[pss,fleqn,embeddedheads]{w-art}
\usepackage{times}
\usepackage{w-thm}
\usepackage[]{graphicx,placeins,subfigure,wrapfig}
\begin{document}
%%    The information for the title page will be placed between
%%    \begin{document} and \maketitle. The order of most entries
%%    is determined by the class file and can not be changed by
%%    rearranging them. The maketitle command follows after the
%%    absract.
%%
%%    Most of the following commands will be completed by the publisher.
%%
\DOIsuffix{theDOIsuffix}
%%
%% issueinfo for header and copyright line

%\Volume{XX}
%\Issue{1}
%\Copyrightissue{01}
%\Month{01}
%\Year{2003}

%%
%%    First and last pagenumber of the article. If the option
%%    'autolastpage' is set (default) the second argument may be left empty.
\pagespan{1}{}
%%
%%    Dates will be filled in by the publisher. The 'reviseddate' and
%%    'dateposted' (Published online) entry may be left empty.
\Receiveddate{zzz} \Reviseddate{zzz} \Accepteddate{zzz}
\Dateposted{zzz} %%
%%    Give a maximum of six PACS code in numerical order.

%% \pretitle{Editor's Choice}

%% We have a short and a long form for the title. The short form
%% (optional argument) goes into the running head.

\title{Diamond thin Film Detectors
for Beam Monitoring Devices}

\author[J. Bol]{Johannes Bol \footnote{Corresponding
     author: e-mail: {\sf hannes.bol@googlemail.com}, Phone: +49\,7247\,82\, 4173,%
     % Fax: +49\,30\,470\,31\,334
     }\inst{1}}
\address[\inst{1}]{Inst. f. exp. Kernphysik, Universit\"at Karlsruhe (TH), Wolfgang-Gaede-Weg 1,
Geb\"aude 30.23, 76131 Karlsruhe, Germany} %%
\author[S. M\"uller]{Steffen M\"uller\inst{1}}
\author[E. Berdermann]{Eleni Berdermann\inst{2}}
\author[W. de Boer]{Wim de Boer\inst{1}}
\author[A. Furgeri]{Alexander Furgeri\inst{1}}
\author[M. Pomorski]{Michal Pomorski\inst{2}}
\author[C. Sander]{Christian Sander\footnote{Now at Inst. f. Experimentalphysik, Univ. Hamburg, 22761 Hamburg, Germany}\inst{1}}
\author[S. Udrea]{Serban Udrea\inst{3}}
\author[D. Varenstov]{Dmitry Varenstov\footnote{Now at GSI, 64291 Darmstadt, Germany} \inst{3}}
\address[\inst{2}]{Gesellschaft f\"ur Schwerionenforschung (GSI), Planckstr. 1, 64291 Darmstadt, Germany}
\address[\inst{3}]{Inst. f. Kernphysik, Techn. Universit\"at Darmstadt, Schlossgartenstrasse 9, 64289 Darmstadt, Germany}

%%
%%    \dedicatory{This is a dedicatory.}
\begin{abstract}
Diamonds offer radiation hard sensors, which can be used directly in primary beams.
Here we report on the use of a polycrystalline  CVD diamond strip sensor as beam
monitor of heavy ion beams with up to $\sim 10^9$ lead ions per bunch. The strips
allow for a determination of the transverse beam profile to a fraction of the pitch
of the strips, while the timing information yields the longitudinal bunch length
with a resolution of the order of a few mm.
\end{abstract}
%% maketitle must follow the abstract.
\maketitle                   % Produces the title.

%% If there is not enough space inside the running head
%% for all authors including the title you may provide
%% the leftmark in one of the following three forms:

\renewcommand{\leftmark}
{J. Bol et al.: Diamond thin Film Detectors for Beam Monitoring
Devices}

\section{Introduction}

\begin{wrapfigure}{r}{0.45\textwidth}
  %\begin{center}
  \includegraphics[width= 0.45 \textwidth]{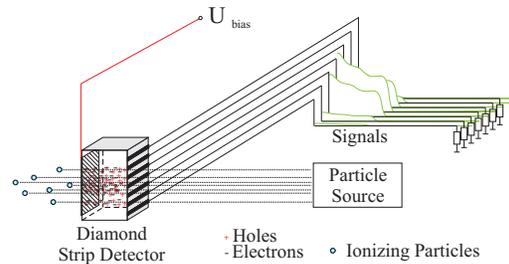}
  \caption{Principle of beam monitoring with diamond detectors.}\label{Principle}
  %\end{center}
\end{wrapfigure}

Diamond is an ideal material for beam monitoring. The idea is simple: A strip
detector, as has been used in high energy physics for a long time, is put directly
in the primary particle beam. (See Fig. \ref{Principle}.) But instead of single
particle tracks, a whole bunch (or a continuous beam) of particles creates a
profile of signals across many strips. Depending on the beam, either the charges
created by a bunch are collected or in case of a continuous beam the current is
measured. Plotting the signal height versus the strip positions results in a
profile which can be used to determine the beam position and width. Using a fast
readout opens the additional possibility of measuring also the time structure of a
beam or a bunch of particles resulting in two dimensional beam intensity
information, (as will be discussed later). Such diamond beam monitors can in
principle replace the usual wire scanners, which measure the transverse profile by
moving a thin wire through the beam and measure the scattered particles outside the
beampipe as function of the wire position. Moving  a diamond strip sensor instead
of a wire into the beam has the advantage that one gets a 2D picture of a {\it
single} bunch instead of a 1D picture averaged over many bunches. This advantage
may be particularly important for electron-positron colliders, where every bunch is
generated, collided and dumped in the beam dump, so it is important to check the
stability of the bunches. For such beam monitors typical resolutions in the $\mu m$
range are required, which can be achieved by small pitch strip detectors. The very
high mobility of charges inside diamond allow for timing resolutions of the order
of 10 ps, so the longitudinal size of the bunch length should be obtainable with  a
precision of the order of a few mm.  In addition, putting orthogonal strips on both
sides of the diamond would lead to 3D profile measurements.

 In this study no
electron beams of this quality were available. Instead the heavy ion plasma beam at
GSI was used, which provides transverse sizes of a few mm and bunch lengths of
about hundred ns. However, it allows to study the principle of a diamond beam
monitor with extreme intensities, so the much more important questions of radiation
damage, stability of metallization, saturation etc. can be studied.

 Using diamond
as detector material has several advantages: a) diamond is much more radiation hard
than silicon \cite{CCE:RD42}. b) diamond has a higher band gap, so it does not need
a depleted diode structure to reduce the leakage current at room temperature.
Consequently it does not have the silicon detector problem of needing depletion
voltages above the breakdown voltage after a high irradiation dose, which increases
the leakage current; c) diamond has a  thermal conductivity five times higher than
the thermal conductivity of copper, which allows for high intensity beams through
the sensor.  The area needed for a beam monitor detector is small, usually 1~cm$^2$
is enough. Therefore the price for a detector grade polycrystalline diamond is
negligible compared with the cost for mechanics and electronics. In this study we
use polycrystalline detector grade CVD diamond films from Element Six \cite{E6} as
sensors with aluminium readout strips sputtered on it using a photolithographic
mask.  The diamond sensor had a thickness of 80~$\mu$m and a size of
1$\times$1~cm$^2$. The metallization was done with aluminum for several reasons.
One is, that the energy loss of the ions in the metallization should be small to
keep the damage low. The others are, that aluminum is easy to bond, easy to process
for the manufacturer of the strips and sticks well  to the diamond sensor. More
details can be found in Ref. \cite{phdthesis}.

\FloatBarrier

\section{Measurements with a heavy ion beam}
The measurements presented here, were performed at the
GSI\footnote{GSI: Gesellschaft f\"ur Schwerionenforschung,
Darmstadt, Germany}. The accelerator facility used for the
measurements consists of two parts. The so called UNILAC, a linear
preaccelerator and the main accelerator SIS18, a synchrotron ring.
There are a few extraction lines from that ring. Our measurements
were performed in a vacuum chamber that could also be filled with
argon.
%(Figure \ref{GSI-chamber} shows a schematic sketch of the chamber while in
%figure \ref{GSI-vacuum-chamber} a photo of the inside can be seen.)
There were two CCD cameras with photocathodes available to acquire images of the
beam passing through the chamber. They detect the light from the recombination of
the excited argon atoms. The accelerator can be used to accelerate almost any kind
of ion from protons to Uranium up to energies of 4715~MeV/u (protons) or 1025~MeV/u
(uranium) \cite{betriebsanweisung, Sis}. In the measurements presented here, a lead
ion ($^{208}Pb^{67+}$) beam was used to generate the signal in the diamond. The
ions had a total kinetic energy of 83.2~GeV corresponding to 400~MeV/u. They were
extracted in bunches of up to 2.5~$\cdot$~$10^9$ ions with a bunch length of
1~$\mu$s. The bunch structure could be chosen with either one or four intensity
maxima. In the latter case the sigma of the bunch  is about 100 ns.

%\begin{figure}
%  \subfigure{\includegraphics[width=0.5 \textwidth]{bilder/gsi-chamber.eps}}
%  \subfigure{\includegraphics[width= 0.4
%  \textwidth]{bilder/gsi-vacuum-chamber.eps}}
%  \caption{Target chamber for the GSI beam test. Drawing (left) and photo (right). }\label{GSI-vacuum-chamber}
%\end{figure}

\begin{figure}
  \hspace{1.5cm}
  \subfigure[Photo]{\includegraphics[width=
  0.38 \textwidth]{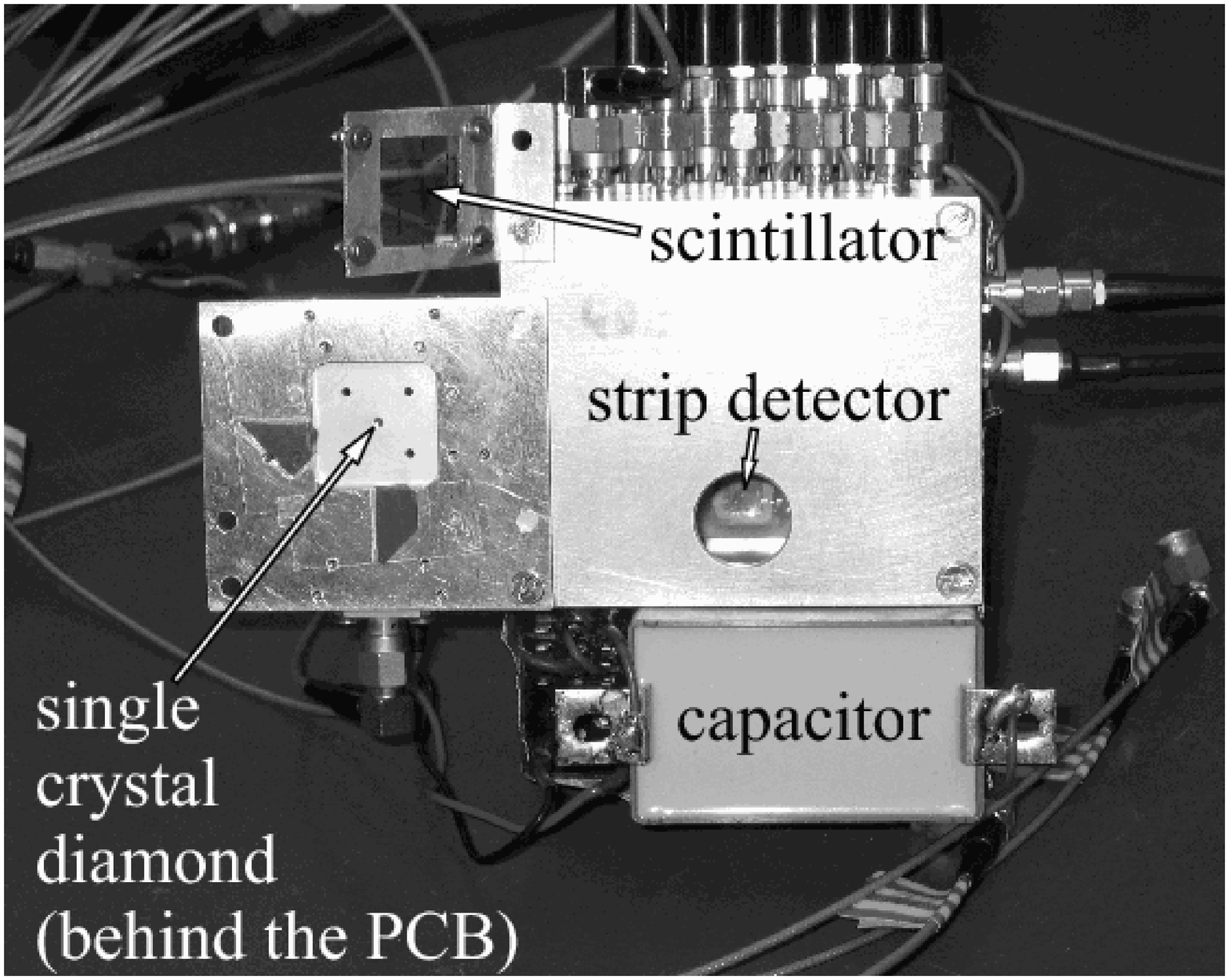}\label{GSI-diamonds}}
  \hspace{1cm}
  \subfigure[Metallization Layout]{\includegraphics[width=0.37
  \textwidth]{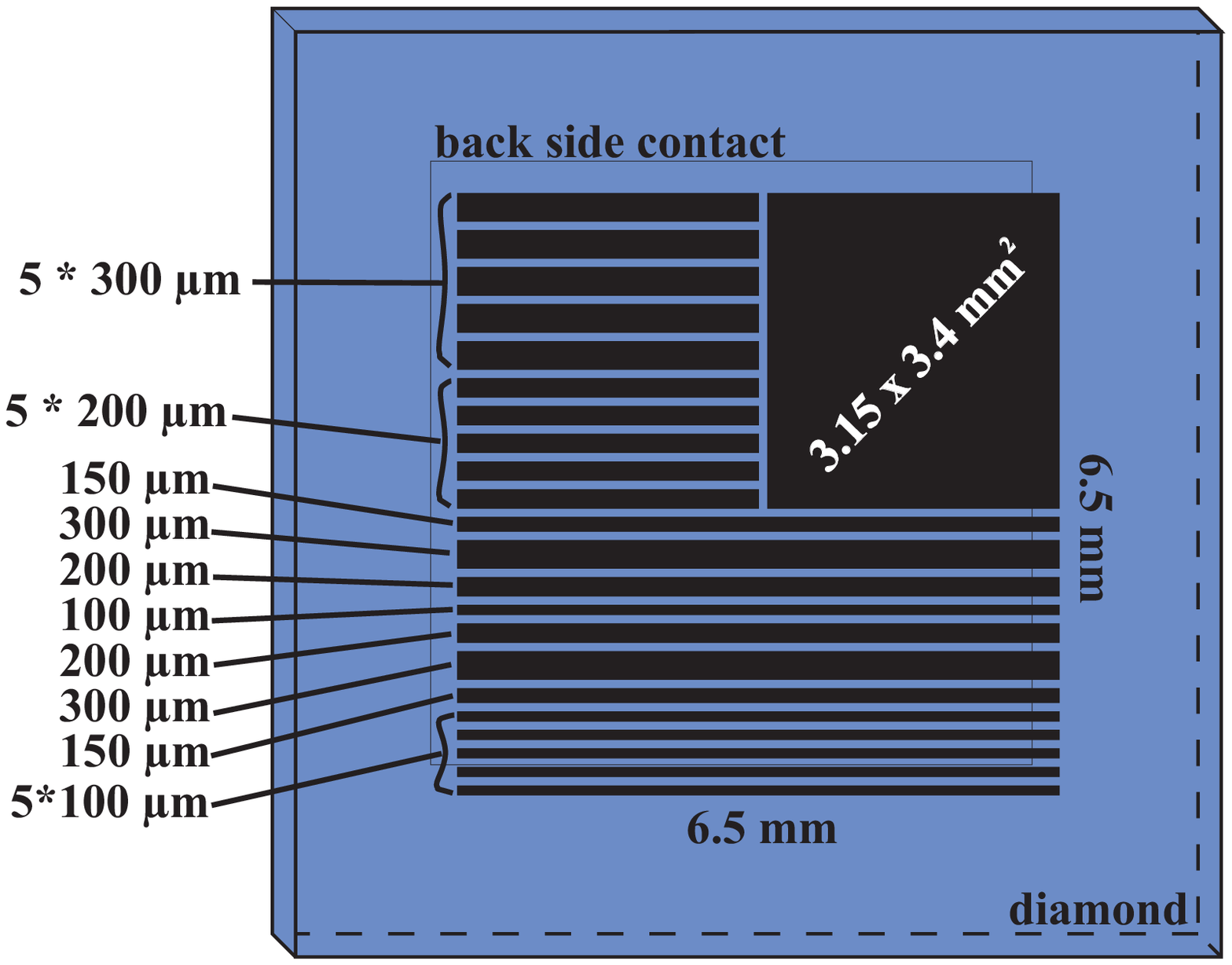}\label{diamondlayout}}
  \caption{The measurement setup  (left) and
  layout of the strips on the polycrystalline CVD diamond sensor (right). The short strips on the
  upper left side and the pad in the upper right corner were bonded together and read out
  via one coaxial cable. The longer strips on the bottom side were all read out separately.
  The results from the single crystal diamond detector are not discussed in this paper.}
\end{figure}

Inside the chamber a computer controlled x-y-z-stage was installed and the diamonds
were mounted on that stage. There were two diamonds, one single pad single crystal
diamond and one polycrystalline strip detector. (See Fig. \ref{GSI-diamonds}.)
Only the results of the polycrystalline strip detector will be presented here. The
layout of the metallization on the strip detector is shown in Fig. \ref{diamondlayout}.

Twelve of the strips (and the larger pad together with the shorter strips) were
connected to a 12 channel GHz-digitizer via the printed circuit board and coaxial cables. Since there
are twelve strips and one pad, one cable had to be switched to read out the pad.
The backplane on the detector, one continuous metallized square, was connected to
the high voltage.

\begin{figure}
  %{r}{0.5\textwidth}
  %\begin{center}
  \hspace{0.5cm}
  \subfigure[Readout]{
  \includegraphics[width=0.35\textwidth]{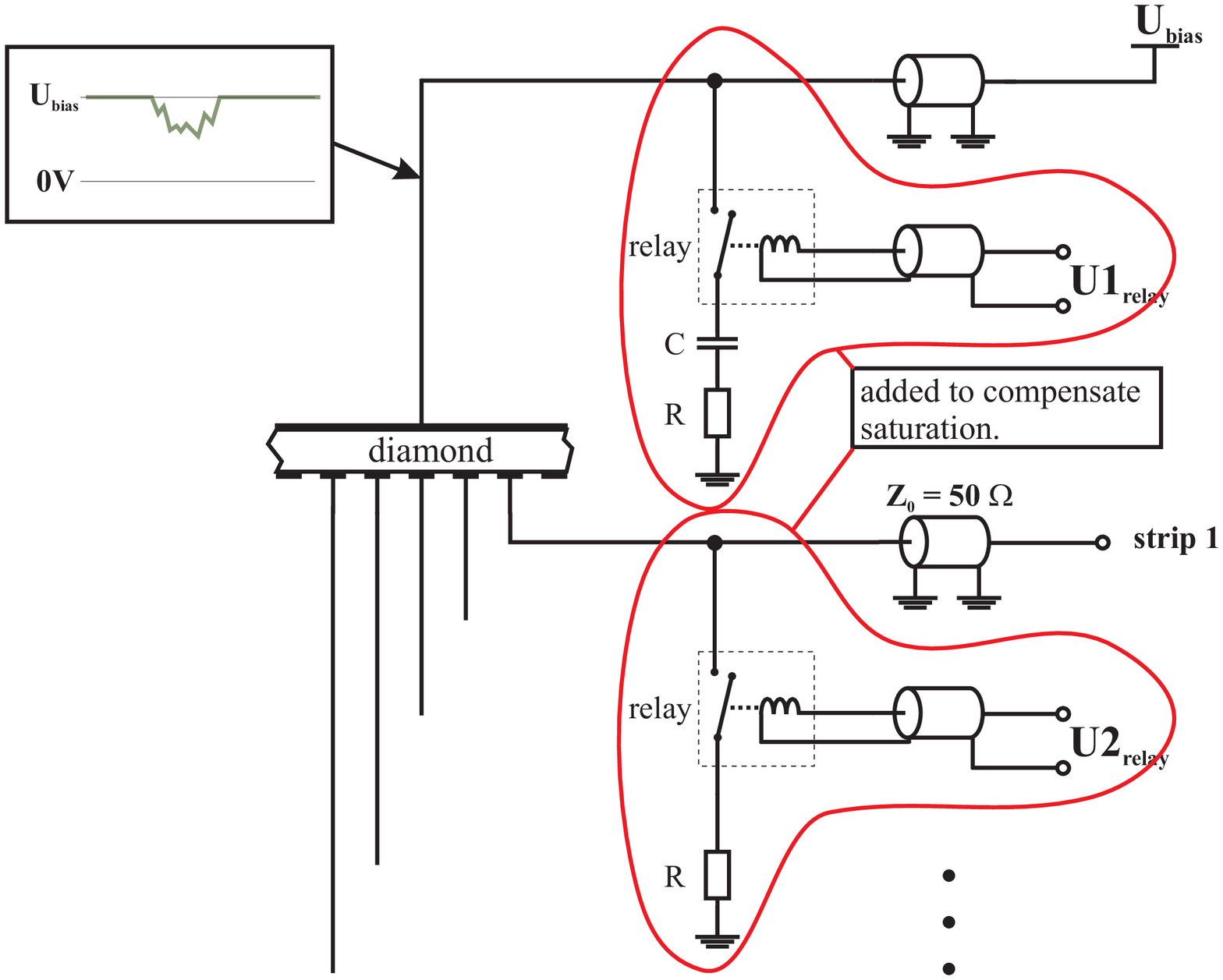}\label{readout-comp}}
  \hspace{1cm}
  \subfigure[Collected Charge for different Impedances]{
  \includegraphics[width=0.45\textwidth]{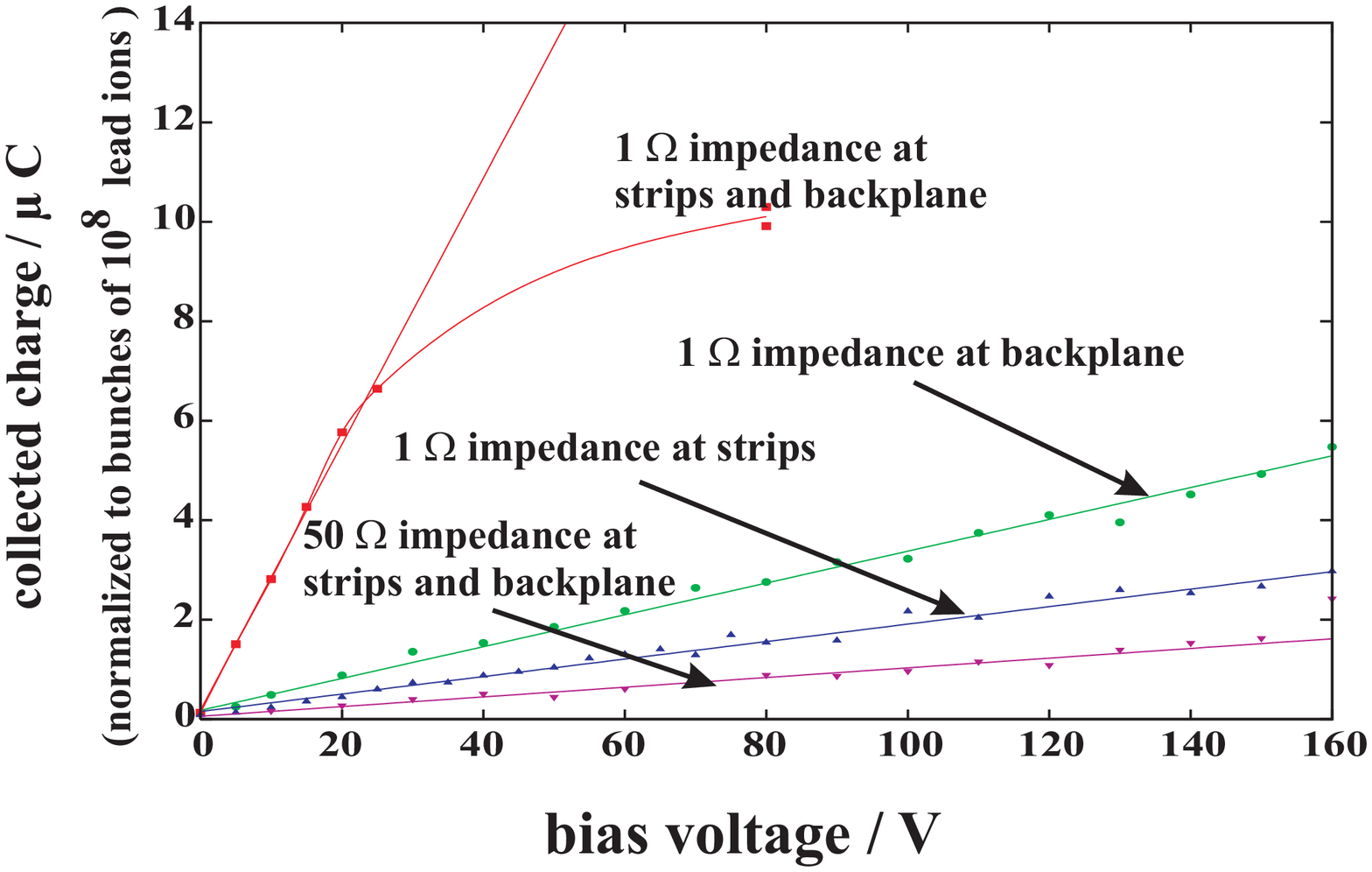}\label{relay-charge}}
  \caption{Simplified schematics of the readout circuit used at the GSI (left).
  The capacitor had a capacitance of 330~nF, the resistors had a value of 1~Ohm.
  The different relay settings
  resulted in different amounts of collected charge(right).}
  %\end{center}
\end{figure}

The readout schematics are shown in Fig. \ref{readout-comp}. In addition to the
simple coaxial cables there are small resistors of 1~$\Omega$ and a capacitor on
the high voltage side to reduce the effect of saturation from beam currents up to
the Ampere range, which implies ionization currents in the kA range for short
bunches. These resistors could be connected and disconnected by relays. The importance
for the measurements will be discussed later.

%\FloatBarrier

\begin{figure}
  \hspace{0.75cm}
  \subfigure[3D-Plot]{\includegraphics[width=0.45
  \textwidth]{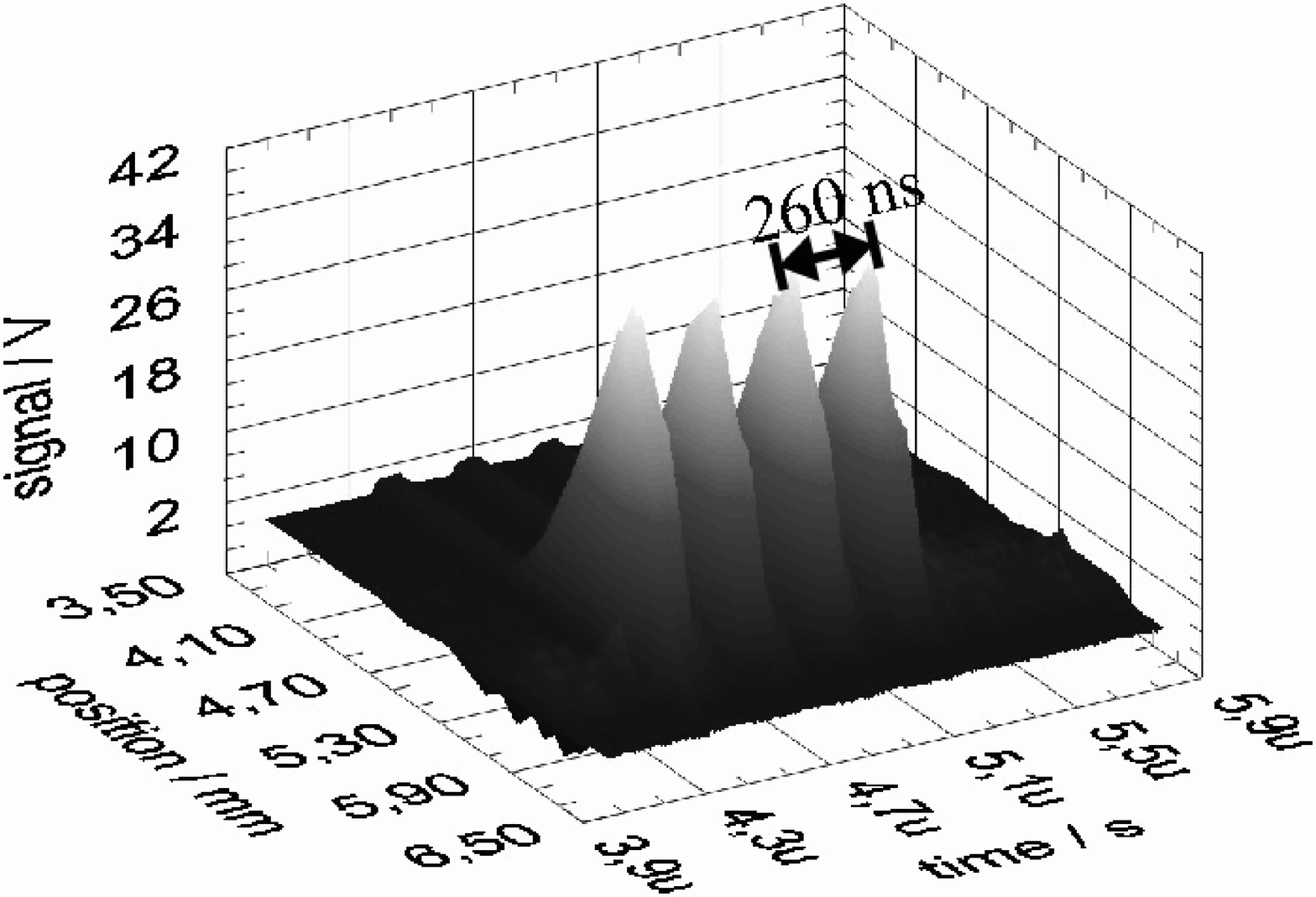}\label{Ch3-LV-quadrupole_3}}
  \subfigure[Signal of the Rogowski Coil]{\includegraphics[width=
  0.45 \textwidth]{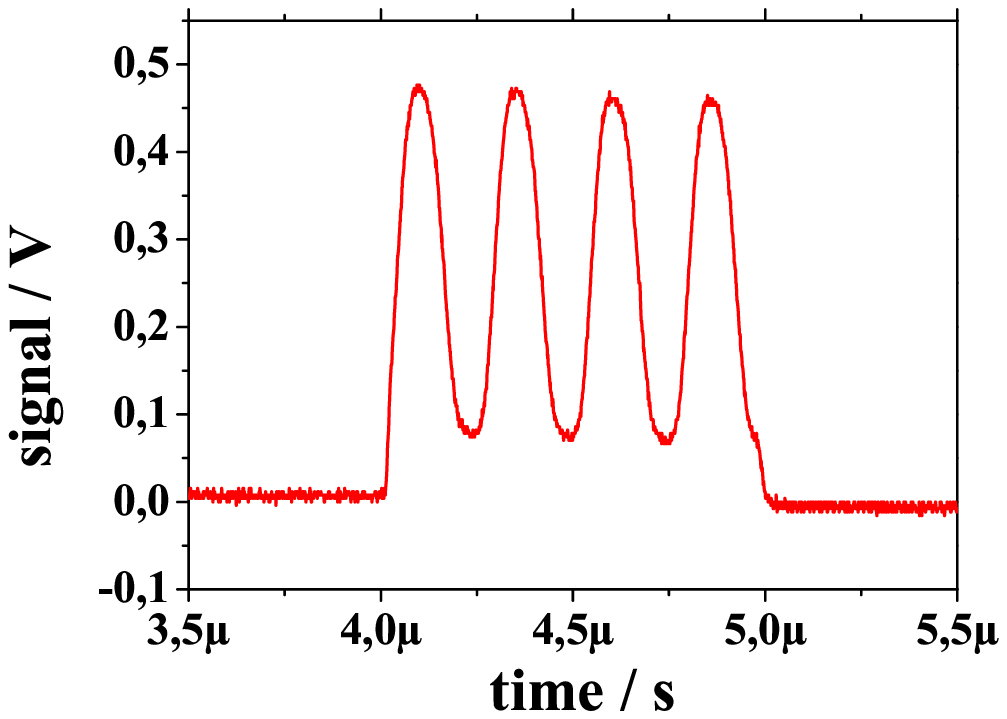}\label{RogowskiGraph1}}
  \caption{Measured data for a single bunch of heavy ions
  with four maxima of the intensity in the bunch shown in a 3D-plot (left) and
  the time structure of the beam as measured with a  pick-up (Rogowsky) coil for comparison (right).}
\end{figure}

An example of a measured beam profile from the lead-ion beam is shown in Fig.
\ref{Ch3-LV-quadrupole_3}. In the left panel the 2D structure of the beam is shown:
the strip information on the left axis gives the transverse profile, while the
timing information (right axis) yields the longitudinal profile. The latter is in
perfect agreement with  the timing structure from a beam pick up coil, plotted in Fig.
\ref{RogowskiGraph1}, thus proving that one can get indeed both the longitudinal
and transverse profile of a single bunch with a diamond beam monitor.

\begin{figure}
  \begin{center}
  \subfigure[Signals of the strips]%
  {\includegraphics[width=0.4\textwidth]{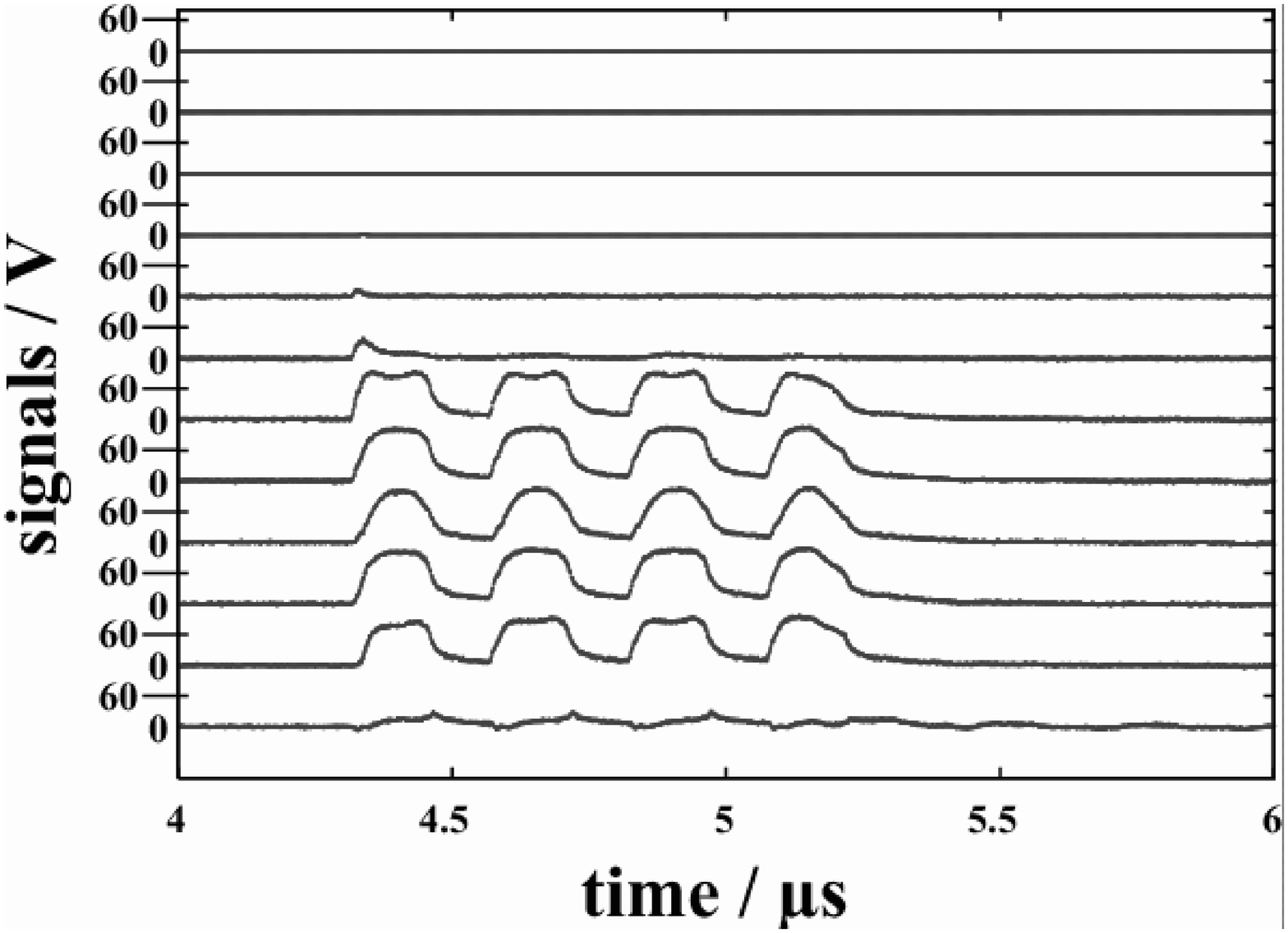}}
  \hspace{1cm}
  \subfigure[Profile]%
  {\includegraphics[width=0.4\textwidth]{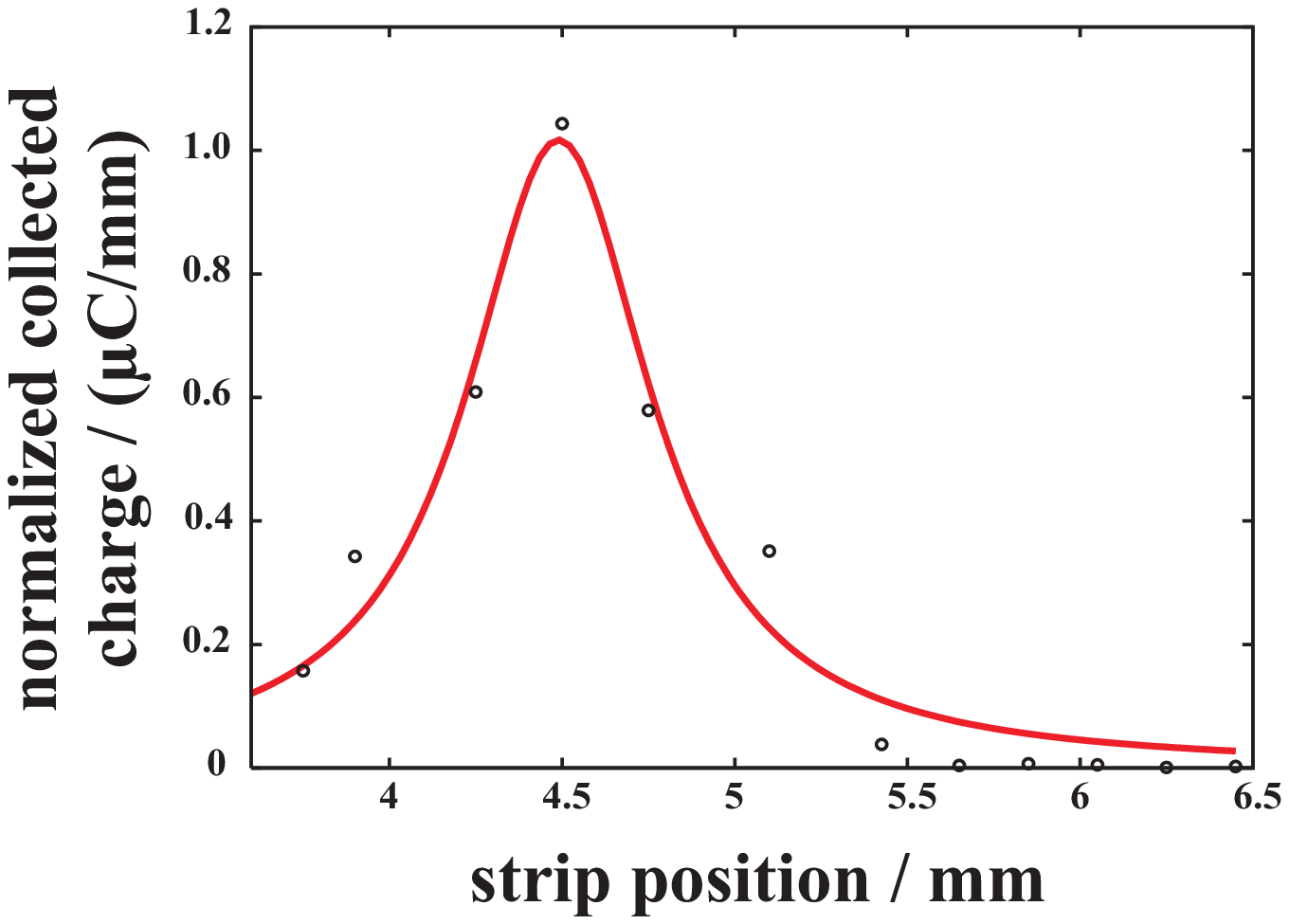}}
  \caption{Oscilloscope picture of all 12 strips (a) and the corresponding beam
  profile (b). A transverse beam profile can be calculated by integrating the
  signals of the four sub bunches of the single strips. The width of the beam can
  obtained by fitting a Lorentz function.}
  \label{sigtoprofile}
  \end{center}
\end{figure}

A quantitative determination of the beam width and beam position requires knowledge
of the collected charge of each strip after correcting for the strip width and
possible non-linear effects from saturation. A raw profile of the 12 digitizer
signals is shown in the left panel of Fig. \ref{sigtoprofile}. Simply normalizing
the signals to the strip width yields the beam profile plotted on the right hand
side. The beam position and width can be determined by simply fitting a Lorentzian
or Gaussian curve.

%\begin{figure}
%  %{r}{0.5\textwidth}
%  \begin{center}
%  \includegraphics[width= 0.5\textwidth]{bilder/RogowskiGraph1.eps}
%  \caption{The beam intensity during a single bunch measured with a Rogowski coil
%  for means of comparison.}
%  \label{RogowskiGraph1}
%  \end{center}
%\end{figure}

%To test the capabilities of the setup measurements were performed
%ith different beam intensities, bias voltages, detector positions and relay
%settings.
%A selection of the measurement results will be
%presented in the following. First of all we tested the effect of the
%relays.
In order to check for non-linear effects measurements were made with several beam
intensities. Ionization currents to the digitizer could be reduced by closing the
switches in Fig. \ref{readout-comp}. High ionization currents to the digitizer
will give an appreciable voltage drop on the 50 $\Omega$ input impedance of the
digitizer, thus reducing the effective voltage on the sensor and the corresponding
charge collection efficiency.
Switching on the resistors divides the short ac-pulse of the ionization current
between these resistors and the
50 $\Omega$ input of the digitizer, thus reducing the voltage drop
on the 50 $\Omega$ readout impedance, and consequently on the detector,
 by a factor 100, if both 1 $\Omega$ resistors are switched on.
In addition, the capacitor helps in stabilizing the
high voltage on the detector when the switches are closed.
Fig. \ref{relay-charge} shows the
effect of the reduced readout impedance.  The total amount of collected charge
increases considerably.
  The observed saturation in that case may be due to the fact
 that all generated charges were
collected for an applied bias voltage of 1~V/$\mu$m.

The disadvantage of the system with the relays are the  relatively long wires needed
to connect the relays. Simulations with SPICE\footnote{SPICE: Simulation Program
with Intgrated Circuit Emphasis} showed that these wires with the relatively large inductance
and corresponding large voltage drop for high current variations in time, as expected for short
bunches, were the reason that the
signals were skewed when the resistors were connected to the strips via the relays.
This problem might be solved in future by connecting the resistors via short
printed circuit connections directly to the strips. Another problem was the
crosstalk between the strips when the resistors were connected. The reason could be
found in a single cable that connected all resistors with ground on the backside.

%\begin{figure}
%  %{r}{0.5\textwidth}
%  \begin{center}
%  \includegraphics[width= 0.5 \textwidth]{bilder/charge2.eps}
%  \caption{Collected charge versus bias voltage for different relay settings.}
%  \label{relay-charge}
%  \end{center}
%\end{figure}

\begin{figure}
%\begin{figure}
  %\begin{center}
  \hspace{0.9cm}
  \subfigure[Detector moved transverse to the beam.]
  {\includegraphics[width=0.4\textwidth]{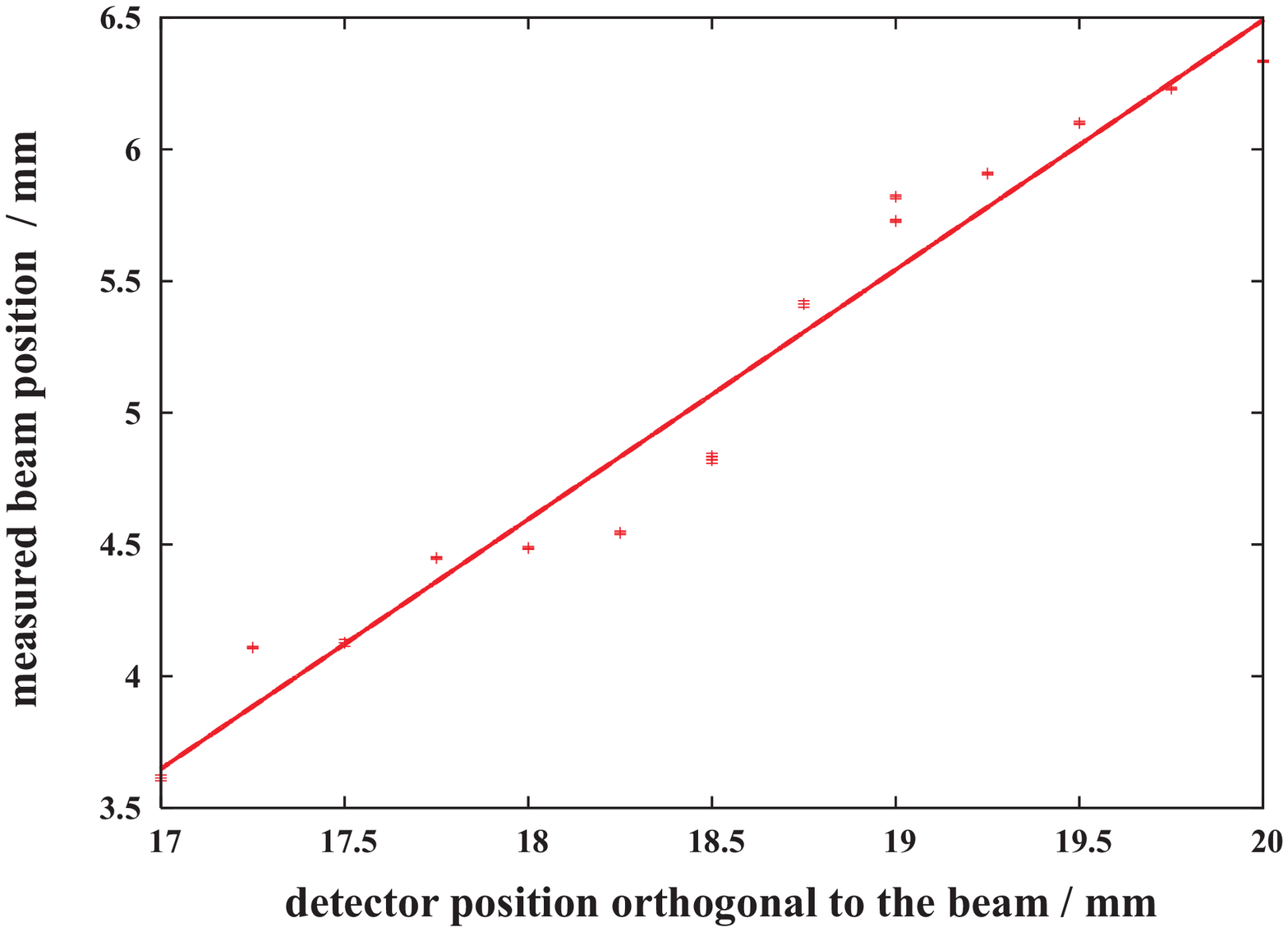}\label{beamposdetpos}}
  \hspace{1cm}
  \subfigure[Detector moved along the beam.]{
  \includegraphics[width=0.4\textwidth]{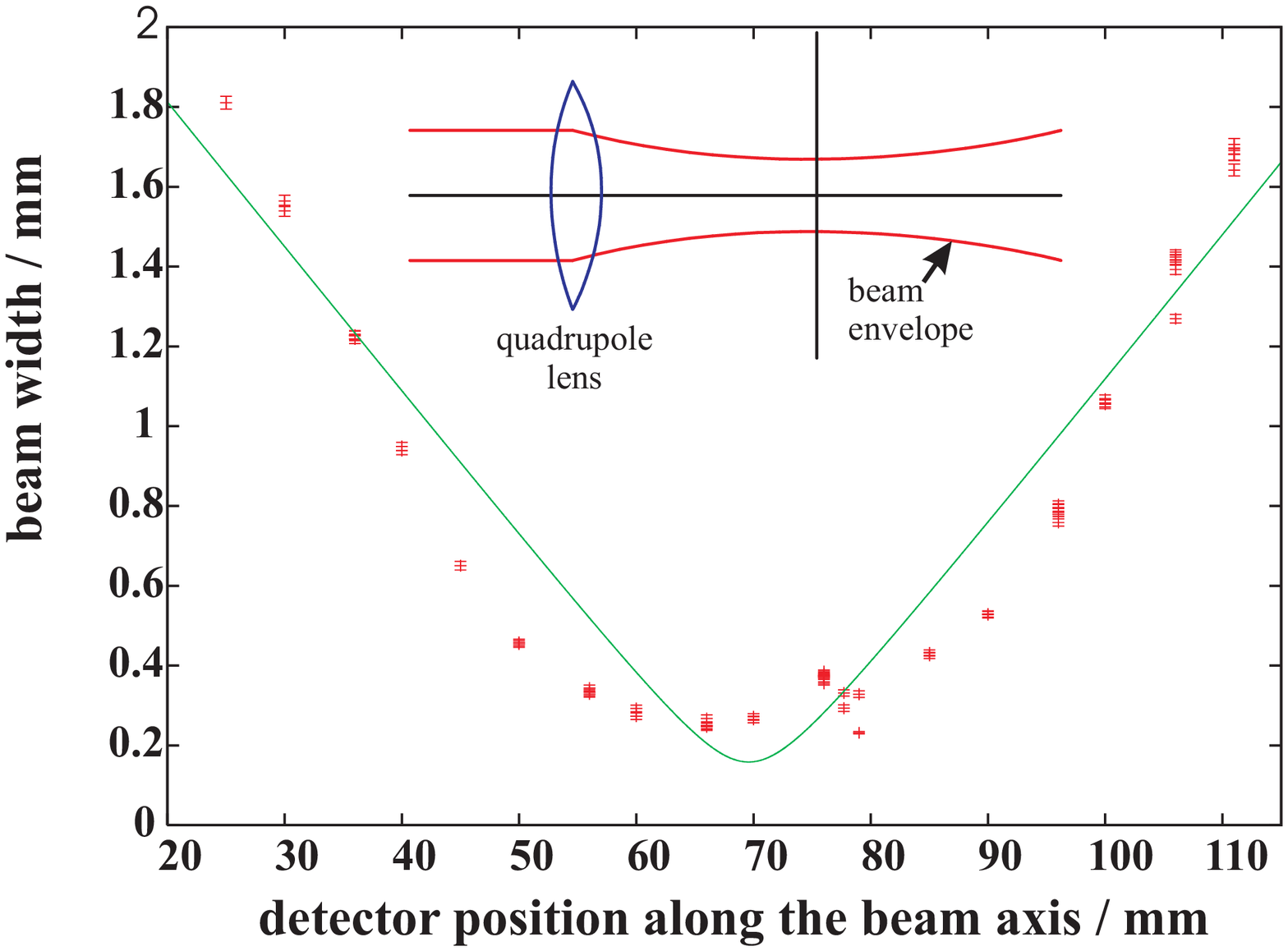}\label{measured-beam-width}}
  \caption{Beam parameters measured at several positions transverse (left) and parallel (right) to the beam
  by the xyz-stage. The curve close to the data points on the right hand side is the expected curve.}
\end{figure}

Another measurement concerns the determination of the beam position. This can be
done by keeping the beam  at a fixed position and moving the  detector orthogonally
across the beam with the x,y-stage. The profile was then determined at known
detector positions and a Lorentz function was fitted to determine the beam
position. The determined beam position is plotted versus the detector position in
Fig. \ref{beamposdetpos}, which shows that there are some positions at
which the measured beam position deviates several hundred micrometers from the
expected positions extracted from the x-y-stage controller. This result is not so
bad taking into account the problems calculating the profiles without worrying
about non-linearities and the large pitch of the strips, which varies from 100~ to 300
$\mu$m (see Fig. \ref{GSI-diamonds}).

Finally some measurements were done to determine the beam width. First of all the
measured profile was compared with a profile determined with a quartz scintillator.
The agreement was good, although not perfect. But it is well known that the
scintillator saturates at the highest intensities in the center of the beam.
Another problem with the scintillator is the fast degradation in the beam. The
second test was a scan of the beam longitudinally along the beam axis. The results
are plotted in Fig. \ref{measured-beam-width} together with a sketch of the
geometry of the beam. The idea of the measurement was to use the fact that the lead
ion beam is focused inside the vacuum chamber by a quadrupole lens directly in
front of it. Therefore there is a relatively steep angle of the beam inside the
vacuum chamber. By moving the detector from one end of the box to the other one
gets a broad variety of different beam widths, as shown in Fig.
\ref{measured-beam-width}. The observed beam width tracks  the expected beam width
quite well. The largest deviations were observed for very
broad bunches and for very narrow bunches close to the focus point.
The first deviations can be explained by the limited width of the area covered by the
strips, while for narrow bunches (in the order of the pitch of the strips) the non-linear effects
and the skewed signals discussed above  leads to uncertainties.

\begin{wrapfigure}{r}{0.45\textwidth}
  \begin{center}
  \vspace{-0.5cm}
  \includegraphics[width= 0.42 \textwidth]{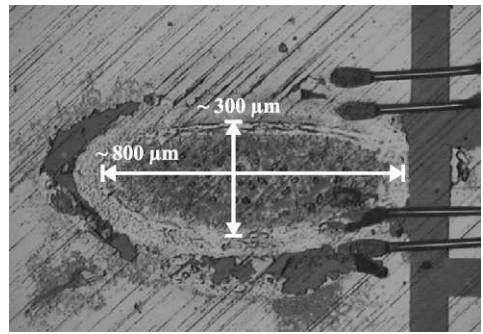}
  \caption{After several shots with the highest intensities on a single spot,
  the imprint from the beam can be clearly seen. However this did not affect the signals.}
  \label{burned-metalization}
  \end{center}
\end{wrapfigure}

As a final measurement  the intensity was increased to the maximum available in order
to test if the diamond could be damaged. The detector was biased and several bunches of up to 2.5 $\cdot
10^9$ lead ions per bunch were fired onto the diamond in a short period of time. A
drop in the signal height was not observed, even after several bunches. The only
damage that could be observed was in the metallization. A microscopic picture of
this damage can be seen in Fig. \ref{burned-metalization}. Although the metal seems
to be evaporated at the edges and in the middle of the beam, the damage appeared to
have no influence on the signal height. But the dimensions of the damage open the
possibility of comparing the measured beam width with the size of the beam.
Obviously the results of measuring the beam width in the
focal point seem to describe the real beam width very well.
%But it is questionable, which beam intensity was required to melt
%the metal, and thus the given dimensions do not necessarily
%represent the beam diameter from the fitted Lorentz function.

\FloatBarrier%
%\vspace{0.5cm}
\section{Conclusions and Prospects}

The presented measurements with a diamond beam monitor show that such a system is
competitive with wire scanners as they are for example foreseen for TESLA
(\cite{TESLA:TDR, wirescanners:Schmidt}). More details on beam monitoring with
diamonds compared to wire scanners can be found in \cite{phdthesis, IEEE:Trans}.
The results prove that such a system gives precise values for the beam position and
width without further measurements of pick up probes or the like. All crosschecks
showed that a high precision can be achieved.

Some improvements can  be made. The main problem is still the large charge
generated in every strip. It is essential to keep the bias voltage applied during
the interaction of the beam to collect all charge, which can be done by a local
capacitor. In addition, a low impedance in the readout chain reduces the voltage
drop on the detector. Additional components close to the detector should be
connected with thick, very short connections in order to have low inductance. This
can be realized with a printed circuit board. A second point is that it would be an
advantage to have a smaller pitch of the strips. Strips on the backside could give
additional information of the beam width and position in an additional dimension.
Smaller strips would also affect the absolute values of the created charge per
channel and therefore reduce problems of voltage drop, thus increasing the dynamic
range. Additionally using a single crystal diamond instead of a polycrystalline
could be interesting for a very narrow beam, as expected in linear colliders. In
very narrow beams the beam dimension could be in the range of the single crystal
grains in the polycrystalline diamond and the variations in the charge collection
efficiency between the grain boundaries could influence the measured profile.
Additional improvements of the system could be the use of still thinner diamonds,
which would suppress the saturation effects.

%Using the experience gained during experiments with heavy ions a
%beam condition monitor based on diamond detectors is being built for
%the LHC\footnote{LHC: Large Hadron Collider} in the
%CMS\footnote{CMS: Compact Muon Solenoid} experiment. It is
%integrated in the large CMS detector to monitor the level of
%scattered particles close to the beam pipe. In case the level
%increases above a certain threshold, a beam dump is initiated. This
%detector is called BCM2. Detailed information about this project can
%be found in \cite{Stemu-dipl-thes, macpherson}.

\begin{acknowledgement}
The authors would like to thank all members - and specially Serban Udrea  and
Dmitry Varentsov - of the plasma physics group of Prof. D.H.H. Hoffman at GSI, who
offered their beam extraction line and experimental setup for our experiments. In
addition we thank the operators of the SIS and Unilac who provided us excellent
experimental conditions.
\end{acknowledgement}

\end{document}